\documentclass[12pt]{iopart}
\usepackage{graphicx}

\begin{document}

\letter{Photoionization of the fullerene ion $C_{60}^+$.}



\author{
R.G. Polozkov$^\ast$, V.K. Ivanov$^\ast$\  and  A.V. Solov'yov$^\dag$
\footnote{On leave from:
A.F. Ioffe Physical-Technical Institute, Russian Academy of Sciences, 
Politechnicheskaya 26,
194021 St. Petersburg, Russia}}
\vspace{16pt} 
\address{$\ast$ St. Petersburg State Polytechnic University,
Politechnicheskaya 29,
195251 St.Petersburg, Russia }
\address{$\dag$ Frankfurt Institute for Advanced Studies,
Johann Wolfgang Goethe University,
Robert-Mayer Str. 10, D-60054 Frankfurt am Main, Germany
}\ead{solovyov@fias.uni-frankfurt.de}
\date{} 

\begin{abstract}

Photoionization cross section of the fullerene ion $C_{60}^+$ has been calculated 
within a single-electron approximation and also by
using a consistent many-body theory accounting for many-electron 
correlations. Electronic wave functions of the ground and excited states have been determined 
within the jellium model based on the local density approximation. 
It is predicted a giant resonance in the photoionization cross section 
at the photon energy $\sim 25$ eV.
It is demonstrated that the resonance profile is much higher and narrower than 
in the case of the neutral fullerene $C_{60}$.

\end{abstract}

Since the discovery till nowadays the fullerene $C_{60}$ attracts a permanent rapt attention     
of investigators. From the physical point of view an interest to this object is associated with its
exotic hollow, but highly stable structure caused by
a significant delocalization of a large number of valence electrons in this molecule. 
These features manifest themselves in a number of interesting physical phenomena occuring in collision
processes involving fullerenes. 
For example, surface plasmon oscillations in the fullerene $C_{60}$ can be excited. The plasmon
excitations have a profound collective nature and they influence significantly 
the formation of the cross sections
in collisions of various kind involving fullerenes  \cite{Plasmons}.
The plasmon excitations in the fullerene $C_{60}$ have been well 
investigated both experimentally and theoretically 
\cite{Bertsch,Bulgac,Yabana,Wendin,Puska,Alasia,Weaver,Hertel,Liebsch,our,our1}. Usually, they manifest themselves  as 
giant  resonances in the excitation spectrum of $C_{60}$.   
In this connection, there arises a question on the possibility of excitation of the
plasmon oscillations in fullerene $C_{60}$ ions of different multiplicity. 
Recently, the photoionization cross section of singly charged 
fullerene ion $C_{60}^+$ has been measured \cite{Scully}. However,  
theoretical calculation of this cross section has not been performed so far.    

In this work, for the first time, we have calculated  the photoionization cross section of 
the fullerene ion $C_{60}^+$  
within the photon energy range from the $C_{60}^+$ ionization threshold up to $40$ eV. 
This energy range has been chosen, because the photoionization cross section for the  neutral fullerene $C_{60}$ 
possesses a very strong giant resonance centered at the photon energy about $20$ eV \cite{Hertel,Liebsch}.
Using the theoretical framework developed earlier by us and 
applied for the description of photoionzation of the neutral fullerene $C_{60}$
\cite{our,our1}, we have determined the location of the giant resonance in the photoionization spectrum
of  $C_{60}^+$ (to be equal $\sim$ $25$ eV) and described it's shape.

The atomic system of units  is used throughout the paper, $m_e =|e|= \hbar =1 $.    

The total photoionization cross section is equal to a sum of partial photoionization 
cross sections for each of the fullerene ion orbitals. The amplitude of ionization of  
each orbital is calculated both within the single-electron approximation and also by taking 
into account many-electron correlations. 
At the first stage, we use the frozen core model
and the local density approximation assuming that there is 
a single electron transition only during the photoionization process. 
Then, the correlations between the transitions from 
different electron states are taken into account within the random phase approximation (RPA).
The RPA amplitudes of photoionization $D_{\nu_2 \nu_1}(\omega)$ 
are found from the matrix form of the RPA equations \cite{Amusia}  
\begin{equation}
D_{\nu_2\nu_1}(\omega)=d_{\nu_2 \nu_1}+
\sum_{\nu_3\nu_4}D_{\nu_2\nu_1}(\omega)\chi_{\nu_4\nu_3}(\omega)V_{\nu_2\nu_1\nu_4\nu_3} 
\label{D}\,.
\end{equation}
Here, the matrix elements $D_{\nu_2 \nu_1}$ are expressed via the single-electron photoionization 
amplitudes $d_{\nu_2 \nu_1}$, the matrix elements of virtual electron-hole excitation $\hat \chi (\omega)$,
and the matrix elements of interelectron interaction $V_{\nu_2\nu_1\nu_4\nu_3}$ \cite{Amusia}. Note
that the excitation spectrum has the Rydberg series of discrete excited states 
because the self-consistent potential of the fullerene ion has the Coulomb asymptotic. Therefore, 
the virtual electron transitions to the discrete excited states of the fullerene ion 
play more important role than in the neutral fullerene $C_{60}$ case. 

The photoionization cross section and the oscillator strengths calculated both within single-electron 
approximation and also by taking into account many-electron corellations have been checked 
on the consistancy with the sum rule.  

The inter-electron interaction  
within the local density approximation  reads as follows \cite{our1}
\begin{equation}
V^{LDA}({\vec r},{\vec r\,'}) = \frac1{|{\vec r}-{\vec r\,'}|} +  
\left.\frac{\partial^2\epsilon_{xc}^{hom}[n]}{\partial n^2}\right |_{n=n_0({\vec r})}\delta ({\vec r} - {\vec r\,'})\,,
\end{equation}                                                               
where $\epsilon_{xc}^{hom}[n]$ is the exchange-correlations energy of homogeneous electron gas 
with density $n$ within the Gunnarson-Lundqvist approximation \cite{Gun} 
\begin{equation}
\epsilon_{xc} = -\frac3{4}\left(\frac9{\pi^2}\right)^{1/3}\frac1{r_s(r)}-
0.0333G\left(r_s\left(\frac{r}{11.4}\right)\right)
\end{equation} 
\begin{equation}
G(x) = (1+x^3)\ln (1+\frac1{x})-x^2+\frac{x}{2}-\frac1{3}\,,
\end{equation} 	
where $r_s(r) = \left(\frac3{4\pi n(r)}\right)^{1/3}$ is the Wigner-Zeith radius, $n_{0}$ is the equilibrium 
density.

Single-electron states of the fullerene ion have been calculated within the  
spherical jellium model with the self-consistent potential \cite{our,our1}  determined
within the local density approximation. 
Analogously to an atom, the single-electron wave functions of $C_{60}^+$ within this model,
are given by  

\begin{equation}
\phi_i\equiv \phi_{n_i,l_i,m_i}({\bf r}) = \frac1{r}P_{n_il_i}(r)Y_{l_im_i}(\theta,\varphi)
\label{spher},
\end{equation}

\noindent where $n_i,l_i,m_i$  is the usual set of quantum numbers for the i-th spherical shell; 
$Y_{lm}(\theta,\varphi)$ is the spherical harmonic; $P_{n_il_i}(r)$ is the radial part of the wave function 
determined in the self-consistent potential of the fullerene ion. 
The number of nodes  of the radial wave function $P_{n_il_i}(r)$ is equal to
$(n_i-l_i-1)$. The main limitation of this method arises from 
the electrons self-interaction. Note that the self-interaction correction does not change the order of 
the electronic energy levels and the value of the total electronic energy, but it alters the value of the
ionization potential very significantly.

As in the case of neutral fullerene, we assumed that four 2s2p
electrons of each carbon atom are delocalized in the fullerene ion $C_{60}^+$. 
The ionic core of the fullerene is formed by fourfold-charged positive carbon ions $C^{4+}$ with
the electronic configuration $1s^2$. The charges of the carbon ions are
averaged over the sphere of radius $R$, being a hypothetic radius of the fullerene ion. 
In this paper we neglect the variation of the fullerene's radius caused by its ionization
and put the radius of the fullerene ion equal to the radius of the neutral fullerene,
i.e. $R = 6.69\, a.u.$ \cite{Haddon}.

For better understanding of the electronic
properties and interatomic bonding in the fullerene molecules,
it is useful to establish the correspondence
of the electronic eigenstates introduced in our
model to the $\sigma$ and $\pi$ orbitals widely used
for modeling the planar graphite surface \cite{our}. 
The ground state configuration for the ion $C_{60}^+$ has been chosen 
in a similar way as for the neutral  fullerene $C_{60}$ \cite{our}. Removing 
one electron from the highest occupied molecular orbital of $C_{60}$, one derives 

\begin{center}{$\underbrace{1s^22p^63d^{10}4f^{14}5g^{18}6h^{22}7i^{26}8k^{30}
9l^{34}10m^{18}}_{\mbox{without nodes, $180 e^-$}}$
$\underbrace{2s^23p^64d^{10}5f^{14}6g^{18}7h^{9}}_{\mbox{one node, $59 e^-$}}$\\
}
\end{center}

Here an atomic notation of electronic configuration $n\,l^N_e$ is used \cite{ll}.

\begin{figure}[t]
\includegraphics[scale=0.5,clip]{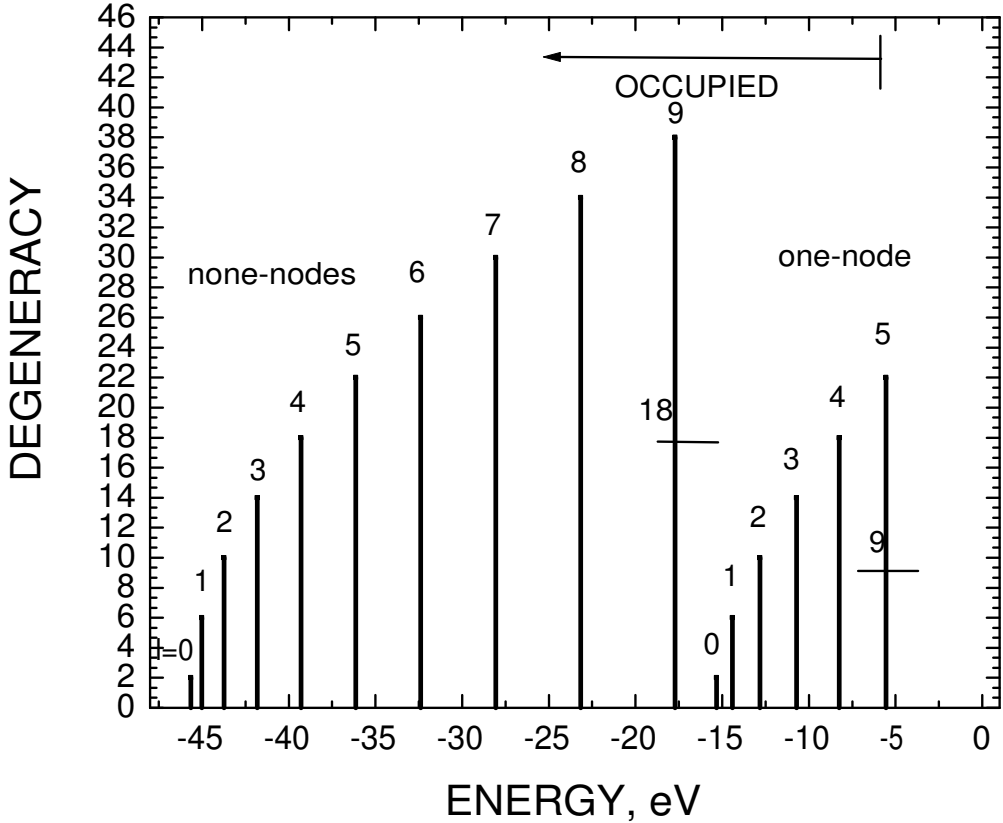}
\caption{
Calculated energy levels spectrum of the
fullerene ion $C_{60}^+$.
The outer none-nodes and one-nodes
orbitals are partially filled.
}
\label{fig1}
\end{figure}

The calculated energy level spectrum is shown in figure 1. The ionization 
potential $5.5$ eV obtained within the LDA  differs 
from the experimental value $11.5$ eV \cite{Seifert}. The origin for this difference  
is discussed above. 

The maximum number of 
electrons in a shell characterized  by the orbital momentum $l$ is equal to $N_l=2(2l+1)$. 
The most outer $\sigma-$ (l=9) and $\pi-$ shells (l=5) in $C_{60}^+$ are open and have
$N_e=18$ and $N_e=9$ electrons respectively, whereas
$N_9=38$ ($\sigma-$ shell) and $N_5=22$ ($\pi-$ shell). 
This means that the total orbital moment $L$ and the total spin $S$
of such a system are non-zero.

It is well known that an accurate calculation of the photoionization cross section 
of an atomic system with open electronic shells is much more complicated than that
for a closed shell atomic system,
because for an open shell system the calculation involves 
a huge number of additional terms which vanish in the case of a closed shell system \cite{Amusia1}.  
In order to avoid this technical problem we have used the averaged term 
approximation. We have assumed that the electrons from the most outer open shells
behave in the photoionization process as if the shells would be closed. This
implies that one can put $L=0$, $S=0$ when calculating the partial 
contribution to the photoionization cross section from these shells and
take into accout their filling ratios $g_l=N_e/2(2l+1)<1$.

It is natural to expect that the averaged term approximation affects the 
photoionization cross section behavior. 
The numerical analysis shows that the  cross section calculated
within the single-electron averaged term approximation does not obey the sum rule.
Therefore, one should introduce appropreate correction
when calculating the photoionization cross section in the
random phase approximation.
Thus, we have input extra coefficients 
decreasing the amplitude of electron-hole excitations from the open shells.
These coefficients have been determined from
the sum rule analysis of the partial photoionization cross sections
of the open fullerene shells performed within the single-electron approximation.
The coefficients enter as numerical factors in the matrix elements of 
operator $\hat \chi (\omega)$ in equation (\ref{D}).
Although this procedure improves the consitence with the sum rule of the photoionization
cross sections calculated within the RPA, it leaves 
the deviation of the RPA photoionization cross sections from the sum rule
on the level of $30$ $\%$. 

\begin{figure}[t]
\includegraphics[scale=0.45,clip]{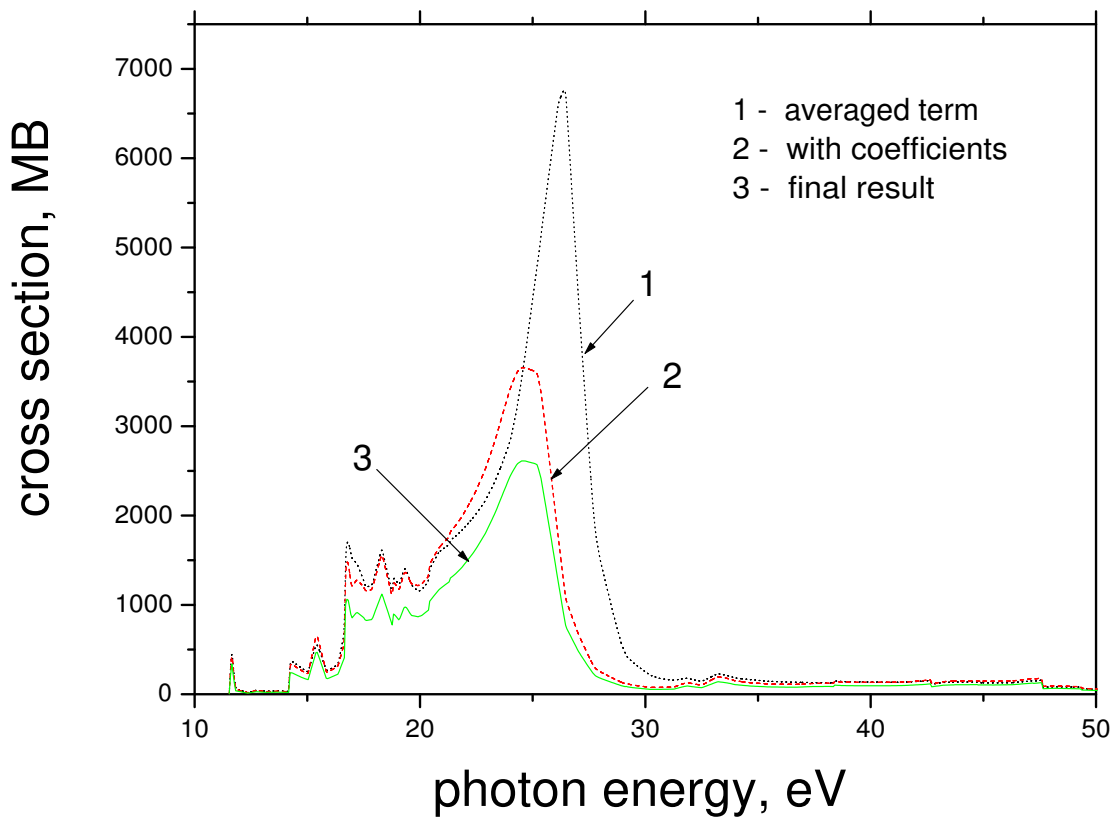}
\caption{
Photoionization cross section of the
fullerene ion $C_{60}^+$ calculated within different approximations.
Dotted line is the result of RPA and the averaged term approximation, 
dashed line is the same as dotted line but obtained with
the optimal values of additional coefficients in the averaged term approximation, 
solid line is the same as dashed line but normalized according to the sum rule.
}
\label{fig2}
\end{figure}

Figure 2 shows
the $C_{60}^+$ photoionization  cross sections calculated by different methods:  RPA and 
the averaged term approximation, 
RPA with extra coefficients,  RPA with extra coefficients 
normalized on the sum rule. The last 
curve is the most accurate result of our calculations of the $C_{60}^+$ photoionization cross section.  
These calculations demonstrate that the giant resonance pattern is not affected much
by the choice of the coefficient in equation (\ref{D}),
but the maximum value of the cross section depends strongly on their choice.

The calculated cross section is compared with theoretical \cite{our1} and experimental 
cross sections \cite{Hertel} 
for the neutral $C_{60}$ in figure 3. In order to elucidate the difference between 
the photoionization cross 
sections for $C_{60}$ and $C_{60}^+$, the $C_{60}^+$ 
photoionization cross section is shifted in figure 3 towards the ionization 
threshold of $C_{60}$. Under the condition 
of coincidence of the ionization thresholds of $C_{60}$ and $C_{60}^+$, the giant resonance 
in the  $C_{60}^+$ photoionization spectrum is
placed at the photon energy $21$ eV. It is well known that the resonance 
energy of the plasmon oscillation  can be easily evaluateded for a spherical fullerene
with the use of the classical Mie formula 

\begin{equation}
\omega _l=\sqrt{\frac{l(l+1)N_e}{(2l+1)R^3}}\,,
\label{Mie}
\end{equation}
where  $\omega _l$ is the frequency of a surface plasmon oscillation with angular momentum 
$l$ and $N_e$ is the number of delocalized electrons.
For the diplole plasmon mode, $l=1$, in the fullerene $C_{60}$ with $R=6.69$ a.u. \cite{Haddon} 
and $N_e = 240$,  one derives $\omega _1 = 20 $ eV.

\begin{figure}[t]
\includegraphics[scale=0.45,clip]{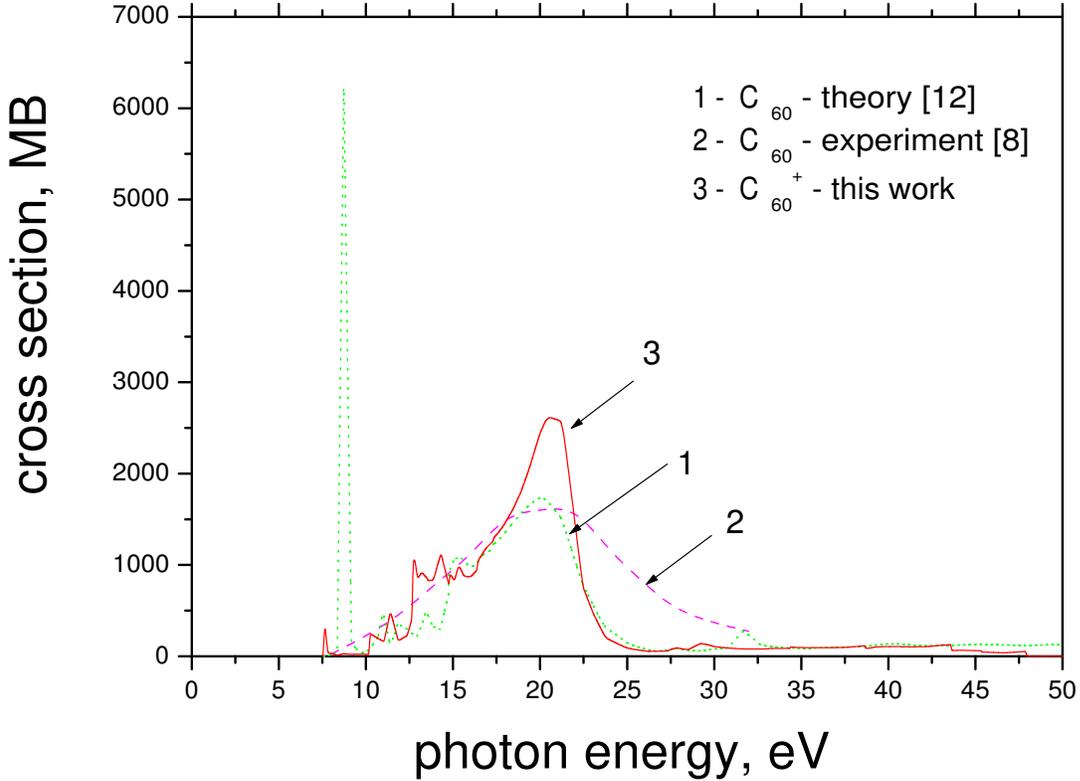}
\caption{
Comparison of the photoionization cross sections for the neutral fullerene $C_{60}$ and the
ion $C_{60}^+$.
Solid line  is the most accurate result of calculation of the photoionization cross section for the
fullerene ion $C_{60}^+$. This curve is shifted on $\Delta=
I_p^{C_{60}}\,-I_p^{C_{60}^+}$, where $I_p$'s are the ionization potentials.
Dashed line shows the experimental photoionization cross section for $C_{60}$ \cite{Hertel},
dotted line is the photoionization cross section for the neutral $C_{60}$  
calculated within our model \cite{our1}.
}
\label{fig3}
\end{figure}

Since we have assumed the same radius for
the fullerene ion as for the neutral fullerene and there is only a small
difference in the number of delocalized electrons, it is natural to expect that
the plasmon resonance energies in the ion and
in the neutral fullerene are close.
Figure 3 illustrates this fact. It shows that nevertheless
there is a slight shift of the 
position of the giant resonance in the $C_{60}^+$ photoionization spectrum, which
might be a result of a slight variation of the radius of the fullerene ion
neglected in our calculations.

Our analysis demonstrates that the giant resonance in the photoionization cross section
for the fullerene ion arises due to the correlation between 
transitions from  $\sigma$$-$ orbitals.
This fact proves the similarity of the nature of
the giant resonances in the cases of the neutral fullerene and the ion. 
Qualitatively, one can explain the origin of 
the giant plasmon resonance by oscillations of the $\sigma-$ electrons density against
the ionic core.

It is also necessary to point out that the giant plasmon resonance in 
the photoionization spectrum of $C_{60}^+$ 
is much stronger as compared with that for $C_{60}$. The similar behaviour is well known 
from the investigation of giant resonances in many-electron atoms and their ions. It
can be explained by enhanced influence of the Coulomb 
core potential on the ground and excited states of valence electrons
in the case of ionic target \cite{Amusia1}.

In conclusion we stress that the method developed in our work allowed us to predict  the  
plasmon resonance pattern in the photoionization cross section of
the fullerene ion $C_{60}^+$. This method can be straightforwardly utilized for the calculation of the 
photoionization cross sections of positive and negative ions of $C_{60}$ of larger multiplicity,
which are of great interest for experimental investigations.

\ \ The authors acknowledge support of this work by 
the INTAS (grant No 03-51-6170) 
and the Russian Foundation for Basic Research {(grant No 03-02-16415-a)}.




\end{document}